# Competing DNS Roots: Creative Destruction or Just Plain Destruction?

TPRC's 29[th] Research Conference on Communication, Information, Cabbages, and Kings. October 2001

Dr. Milton L. Mueller
Graduate Program in Telecommunications and Network Management
Syracuse University School of Information Studies

## Abstract

The Internet Domain Name System (DNS) is a hierarchical name space that enables the assignment of unique, mnemonic identifiers to Internet hosts and the consistent mapping of these names to IP addresses. The root of the domain name system is the top of the hierarchy and is currently managed by a quasi-private centralized regulatory authority, the Internet Corporation for Assigned Names and Numbers (ICANN). This paper identifies and discusses the economic and policy issues raised by competing DNS roots. The paper provides a precise definition of root-competition and shows that multiple roots are a species of standards competition, in which network externalities play a major role. The paper performs a structural analysis of the different forms that competing DNS roots can take and their effects on end-user compatibility. It then explores the policy implications of the various forms of competition.

The thesis of the paper is that root competition is caused by a severe disjunction between the demand for and supply of top-level domain names. ICANN has authorized a tiny number of new top-level domains (7) and subjected their operators to excruciatingly slow and expensive contractual negotiations. The growth of alternate DNS roots is an attempt to bypass that bottleneck. The paper arrives at the policy conclusion that competition among DNS roots should be permitted and is a healthy outlet for inefficiency or abuses of power by the dominant root administrator.

# 1. Introduction and Overview

Three years after the creation of the Internet Corporation for Assigned Names and Numbers (ICANN), management of the Domain Name System (DNS) continues to generate controversy. One of the most intense and significant policy controversies associated with ICANN is the problem of competing DNS roots. Within Internet circles, the problem of competing roots has taken on the characteristics of a religious war. Internet "Catholics," a collection of veteran techies acknowledging Vint Cerf as their Pope and the late Jon Postel as the Messiah, demand allegiance to The One True Universal Root. [IAB, 2000; Lynn, 2001] Internet Protestants, with elected ICANN Board member Karl Auerbach assuming the role of Luther, insist on the freedom to create alternate roots – but just as Luther's challenge led to the proliferation of hundreds of sects and denominations, so (some fear) will the proliferation of DNS roots fragment the Internet.

My goal in this paper is to identify and discuss the economic and policy issues raised by competing DNS roots. I begin with the observation that multiple roots are a species of standards competition, in which network externalities play a huge role. I then perform a structural analysis of the different forms competition among DNS roots can take, and analyze their compatibility effects and policy implications.

In the established domain name market, the value that can be added by competing DNS roots seems to be small relative to the compatibility and fragmentation risks.[1] So why are we getting competing DNS roots? The answer is clear: root competition is caused by ICANN's extremely restrictive and slow addition of new top-level domains to the domain name system. There are no technical constraints preventing the addition of thousands of new top-level names to the DNS root. There are many willing suppliers of new top-level domains, and many members of the public are willing to pay to register them. The ICANN regime, however, has deliberately maintained an extreme scarcity in the supply of top-level domain names. It has authorized a tiny number of new top-level domains (7) and subjected their operators to excruciatingly slow and expensive contractual negotiations. The growth of alternate DNS roots is an attempt to bypass that bottleneck.

More fundamentally, competing roots are about competing sources of authority over the Internet. The basic issue they raise is: Should the direction of the DNS market be driven by technology and the market, or by a quasi-governmental collective action via ICANN?

The policy conclusion I draw is that competition among DNS roots should be permitted and is a healthy outlet for inefficiency or abuses of power by the dominant root. Such competition does not threaten the universality of the Internet interconnection, because both suppliers and end users have powerful economic incentives to remain compatible and connected with each other. Indeed, the deck is stacked so heavily in favor of an established root that if an alternate root achieves critical mass and threatens the dominance of the existing root, it can only happen because the existing root is doing something seriously wrong. ICANN itself claims to be based on an "Internet community consensus." If so, it has nothing to fear from competing roots, because – given the importance of network externalities in this market – a sustainable alternate root could only arise from a clear lack of consensus support for ICANN policies among a critical mass of suppliers and consumers.

---

[1] The domain name registration market is valued at about US$ 2 billion annually. For the ordinary business or nonprofit organization holding one or two domain names, the cost of the domain name ($20-50 per year) is extremely small compared to the other costs associated with an Internet presence (servers, labor, ISP service, etc.).

A more difficult and complex issue to be faced is how much interconnection should be implemented between a dominant root and alternate systems. If ICANN continues to stifle the market for new top-level domains, this issue will have to be faced. In this paper, I sketch out some of the issues involved in interconnecting roots, but do not make any detailed or specific proposals.

## 2. The DNS Name space

The DNS is a distributed database protocol that allows end users around the world to coordinate the assignment of unique names to computers. The protocol also coordinates the translation of those names into the numerical IP addresses that guide the movement of packets across the Internet.

The DNS name space is a hierarchy. (Figure 1) At the top of the hierarchy is an unnamed root. The root is defined by the contents of the *root zone file,* a list of top-level domain name assignments with pointers to primary and secondary name servers for each top-level domain. Currently, the root server system consists of 13 name servers placed in various parts of the world. The server where the root zone file is first loaded is considered authoritative; the others merely copy its contents. The additional servers make the root zone file available more rapidly to spatially distributed users, and provide redundancy in case some root servers lose connectivity or crash.

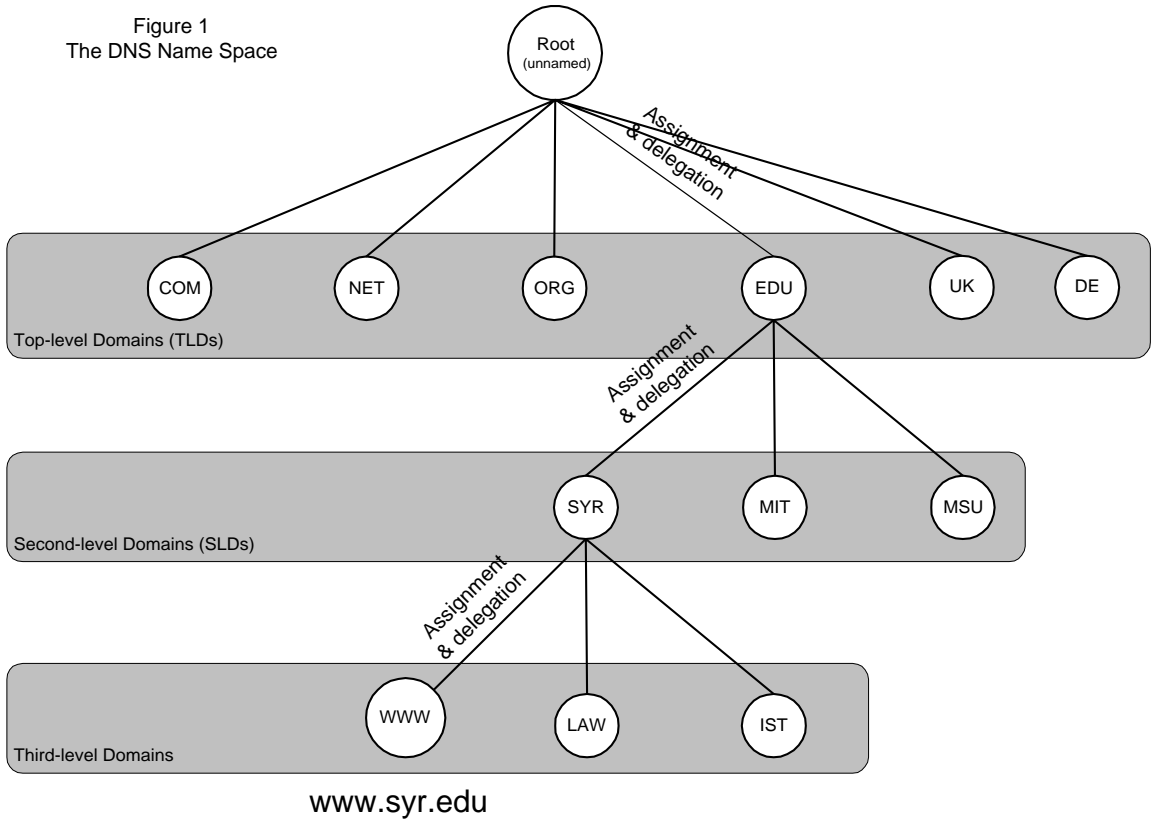

Figure 1
The DNS Name Space

www.syr.edu

Technically, the most important thing about the DNS root is that it provides a single, and therefore globally consistent, starting point for the resolution of domain names. As long as all the world's name servers reference the same data about the contents of the root zone, the picture of the name resolution hierarchy in one part of the world will continue to match closely the picture in any other part of the world.

Aside from its technical significance, administration of the DNS root is important politically and economically. Whoever controls the root zone file also determines who is delegated authority over top-level domains. If top-level domain assignments are economically valuable, and many people believe that they are, then the decision about who gets one and who doesn't gives the assignment authority a great deal of leverage over applicants. Monopoly control of top-level domain name assignments can thus provide the power needed to regulate registry policies, second-level domain name assignments and other aspects of Internet use.

## 3. Competing Roots: Definition and Analysis

Competition at the root level means competition for the *right to define the contents of the root zone file.* The root zone file is the list of recognized top-level domain (TLD) strings (such as COM, NET, ORG, DE, MX, etc.) with pointers to the IP addresses of the top-level name servers capable of resolving names registered under that TLD. When implementing the DNS protocol, anyone with control over the configuration of a DNS client or name server has the ability to choose which root they query for resolution of domain names.[2] Competition, therefore, means that organizations compete to persuade Internet service providers, domain name server operators, and end users to direct root-level DNS queries to them, so that *their* definition of the content of the root zone is recognized and accepted by the rest of the Internet.

Competing roots are a form of standards competition. In standards competition, user choices are affected by the value of compatibility with other users, not just by the technical and economic features of the product or service itself. This feature of demand is called the *network externality* in economic jargon. The value of a system or service to its users tends to increase as other users adopt the same system or service. A more precise definition characterizes them as demand-side economies of scope that arise from the creation of complementary relationships among the components of a system. (Economides, 1996) Network externalities can be realized in one of two basic ways: either all users converge on a single system, or gateway technologies can be developed that interconnect the separate systems or make them compatible.

There are many historical examples of competition based on compatibility, such as studies of competing railroad gauges (Friedlander, 1995), alternate electric power grid standards (Bunn and David, 1988), separate telegraph systems (Brock, 1981), non-interconnected telephone networks (Mueller 1997), and alternate broadcast standards (Farrell and Shapiro, 1992). All have shown that the need for compatibility among multiple users led to convergence on a single standard or network, or in interconnection arrangements among formerly separate systems. In some cases, political and economic factors can limit the scope of convergence (e.g., common broadcast standards were limited to world regions rather than becoming global in scope). Universal compatibility is not always necessary or desirable.

However, in the case of DNS the need for unique domain name assignments and universal resolution of names creates strong network externalities in the selection of a DNS root. If all ISPs and users rely on the same public name space – the same delegation hierarchy – it is likely that all name assignments will be unique, and one can be reasonably confident that one's domain name can be resolved by any name server in the world. Thus, a public name space is vastly more valuable as a tool for internetworking if all other users also rely on it, or coordinate with it. Network administrators thus have a strong tendency to converge on a single DNS root.

Alternate roots face a serious chicken and egg problem. The domain name registrations they sell have little value to an individual user unless many other users utilize the same root zone file information to resolve names. But no one has much of an incentive to point at an alternate root zone when they have so few users. As long as other people don't use the same root zone file, the names from an alternate root will be incompatible with other users' implementation of DNS. That is, other users will be unable to resolve the name, making communication using that name impossible.

---

[2] In actual practice, default values are set in the implementation software, which for 90% of the world's computers is a software known as BIND.

Network externalities are really the *only* barrier to all-out competition over the right to define the root zone file. A root server system is just a name server at the top of the DNS hierarchy. There are hundreds of thousands of name servers being operated by various organizations on the Internet. In principle, any one of them could set up a public name space, assign top-level domain names to registries, and either resolve the names or point to other name servers that resolve them at lower levels of the hierarchy. The catch, however, is that names in an alternate space are not worth much unless a critical mass of name servers on the Internet recognize the alternate root and point their name servers at it.

A major Internet industry player could quickly overcome the critical mass problem. An America Online, a Microsoft, a major ISP such as MCI WorldCom, all possess the economic and technical clout to establish a viable alternate DNS root should they choose to do so. If the producers of Internet browsers, for example, pre-configured their resolvers to point to a new root with an alternate root zone file that included or was compatible with the legacy root zone, millions of users could be switched to an alternate root. It is also possible that a national government with a large population that communicated predominantly with itself could establish an alternate root zone file and require, either through persuasion or regulation, national ISPs to point at it. Indeed, the Peoples Republic of China is offering new top-level domains based on Chinese characters on an experimental basis. The whole transition to multilingual names is creating numerous opportunities for establishing alternate roots.

## Compatibility Structures of Competing Roots

The debate over competing roots can be further clarified by modeling various structures that competition might take and their impact on end user compatibility. If there are two DNS roots – say, the ICANN-Department of Commerce-managed root (Root I) and a competing root (Root C) – competition can take one of three forms [Auerbach, 2000]:

- Type 1: The zone files of Root-I and Root-C have identical contents due to mutual recognition and coordination.
- Type 2: Root-C has more top-level domains than does Root-I but for those TLDs in common, the contents of the root zone are identical.
- Type 3: Root-C and Root-I contain at least one, and possibly many more, conflicting assignments of top-level domains. In other words, each root administrator has assigned the same top-level character string, but to different organizations. The name servers operated by those organizations contain different zone file contents for the top-level domain in question; i.e., they resolve to different IP addresses.

In Type 1 competition, the identical content of the root zone file is achieved through coordination. That is, Root-I agrees to recognize and name serve top-level domains created by Root-C, and vice versa. Both roots agree to avoid making top-level assignments that conflict with TLDs in the other root. The economic effect is identical to an interconnection agreement in telecommunications. The method of coordination can be either social-institutional in nature (e.g., meetings and procedures for assigning names and resolving conflicting claims) or technical (i.e., some kind of protocol that keeps assignments within the roots consistent, or adds information to inconsistent assignments to make them unique).

In Type 2 competition, the competing root adds capabilities (new TLDs) that do not exist in the other root. This is the form of competing roots that we are now experiencing. New.net, ORSC, and other alternate roots all carry the ICANN root zone and do not make assignments that conflict with it. They add new TLDs to the content of the ICANN root zone. This method of competition preserves universal compatibility in all the existing domains, and for all users of Root C, but creates incompatibility for users of Root I. Users of Root-I will *not* be able to resolve names that are unique to Root-C. For example, if Root C adds a top-level domain ".*new*" to its root, its own customers will be able to fully utilize domain names ending in .*new*, as well as all the Root-I top-level domains. But users of Root-I will not be able to resolve domain names ending in .*new*. The compatibility relationships look like this:

**Table 1: Type 2 Competition – Compatibility Relations**

| | Origin of domain name query | |
|---|---|---|
| *Origin of domain name assignment* | **Users of Root-I** | **Users of Root-C** |
| **Root-I** | Compatible | Compatible |
| **Root-C** | Incompatible | Compatible |

In Type 3 competition (Table 2), the root zones are uncoordinated. No user of either root can be sure that they have a unique domain name. Users have a high risk of experiencing incompatibility when they interact with each other (unless of course they all converge on the same root). There are very powerful economic incentives limiting the possibility of such a scenario. Over the medium and long terms, it is difficult to understand why customers would continue to buy, or how suppliers could sustain a business selling, domain names that did not work a significant amount of time, and were perpetually at risk of being incompatible with a significant portion of the Internet. Both ISPs and their customers have very strong incentives to maximize compatibility, and could be expected to converge on a common root or to coordinate the alternate roots.

**Table 2: Type 3 Competition – Compatibility Relations**

| | Origin of domain name query | |
|---|---|---|
| *Origin of domain name assignment* | **Users of Root-I** | **Users of Root-C** |
| **Root-I** | Compatible | Incompatible |
| **Root-C** | Incompatible | Compatible |

## Policy Issues

The three types of root competition enumerated above are useful categories for analyzing the policy issues regarding competing DNS roots.

While Type 1 (interconnection or mutual recognition) might be optimal from the standpoint of the market participants' ability to realize the benefits of network effects, it would severely undermine ICANN's ability to impose public policy and regulation upon the domain name system. In effect, an interconnection agreement with an alternate root would give providers a way to operate new top-level domains without going through ICANN's centralized contracting process. This would in turn provide end users with a choice of policies – if they did not like ICANN's domain name dispute resolution policy, for example, registering a domain name in the alternate root system might provide a way to avoid it. In order to be able to impose policy obligations on DNS providers and users, ICANN must maintain the exclusivity of its root. Whether ICANN's capacity to enforce policy is something worth preserving of course depends on one's opinion of ICANN and its policies. It may well be that both ICANN and the public would be well served by some competition that gave Internet users a choice of policies. But ICANN itself has no incentive to interconnect with an alternate root until and unless its number of users becomes too large to ignore.

Even if ICANN was inclined to recognize the existence of an alternate root, it would still have to establish some thresholds and recognition criteria. It is easy to establish an alternate root with new TLDs if one needn't worry about having real customers. Moreover, indiscriminate recognition of any alternate root that came along would simply encourage new alternate roots to stake indiscriminate claims on top-level domain strings. Some kind of criteria for distinguishing between "real" and "unreal" or "serious" and "silly" alternate roots would have to be established.

A policy white paper put forward by New.net proposes a general solution to this problem. [New.net, 2001, p. 4-5] It analogizes the addition of new TLDs to the root to the decision by local cable television operators to add programming channels to their menu of offerings. A new TLD would seek to gain support among ISPs and users in a similar way. As an alternative to ICANN's highly political and slow TLD-licensing process, New.net proposes that ICANN encourage private companies to develop top-level domains in the market and set some minimal technical standards and a "minimum number of domain names being used by disparate users." When those thresholds are met the TLD is added to the public root. One problem with this approach is that it would not easily allow for the creation of TLDs oriented toward private companies that wanted to internalize their DNS service (e.g., a *.aol* or a *.ford*), or TLDs that served relatively small communities of users (e.g., there are less than 100 registrants in *.int*)

Some analysts have argued that competing roots have the potential to "seriously disrupt the Internet" because of the dissemination of inconsistent data throughout the DNS [Lynn, 2001; Crispin, 2001]. That argument, however, applies *only* to Type 3 competition; i.e., when there is no coordination among the separate roots' name assignments whatsoever and numerous conflicting TLD assignments exist. Type 2 competition does not create such problems, because it retains the consistency of all assignments under TLDs in Root-I, and simply adds new, non-conflicting TLDs to the Root-C zone file. The deliberately deceptive equation of Type 3 and Type 2 competition is one of the more egregious rhetorical tactics used in the DNS wars.

The real policy issue in Type 2 competition comes when the dominant root wants to add new TLDs. At that point it must decide whether or not to assign top-level domains that collide with (i.e., duplicate) those already assigned by the alternate root. For example, if ICANN ever decides to add new top-level domains, it will find that the *.hola* string is already in use by many customers of New.net. If ICANN avoids assigning *.hola* to any registry licensed through its own process for that reason, it is tacitly recognizing New.net, a kind of halfway house toward a Type 1 situation. Of course, avoidance of conflicting assignments is not the same as mutual recognition and coordination – the alternate root's TLDs still would not be listed in the ICANN root zone files. But conflict-avoidance would constrain ICANN's choices of TLD strings, and make it easier for the competing root to survive.

On the other hand, if ICANN assigns TLDs already in public use by another root, it is initiating Type 3 competition. The effect is to create bilateral incompatibilities in the market for domain names. Because of the powerful network externalities, end user reaction to those incompatibilities would in most cases tip the market toward one of the competing roots (in this case, probably the dominant one). For example, if ICANN chose to assign *.hola* to someone other than New.net, it could transform all holders of New.net's *.hola* names into "orphans" and render worthless their prior investments in the names.[3] Such an exercise of market power by the dominant root (ICANN) raises a serious competition policy issue. If it was not illegal for New.net to enter the market in the first place, should it be legal for ICANN to use its dominance of the DNS root-server system to put them out of business? In this scenario, it is important to note that it would be ICANN, rather than New.net, that initiated the incompatibility.

One suggestion for resolving such conflicts is that a right of first use be recognized, similar to the establishment of rights over trademarks. Here again, workable criteria and thresholds must be established.

One option that does not seem feasible is to make alternative roots illegal. Any attempt to require all users to point to an "authoritative" root raises a number of serious policy problems. First, it violates a foundational principle of the Internet, which is that the Internet is a collection of private networks that choose to interconnect (or not) with each other. Second, such "authority" to be effective would have to be global and legally enforceable, which means that it would have to come from an international agreement among governments. Furthermore, any attempt to designate and make mandatory an authoritative root would risk freezing Internet naming technology in its tracks. It would not be easy or unambiguous to define what "pointing to the authoritative root" means in legal terms; such a definition would have to careful not to outlaw technologies that might enhance the DNS, such as content distribution networks, keyword systems, or private name spaces. Nor would it be wise to attempt to prohibit efforts to bypass DNS altogether with a radical new naming or addressing technology.

---

[3] (By the same token, if New.net were well-established enough and had lots of functional registrations under *.hola*, it is possible that the ICANN delegee could fail to attract customers because of the conflict.)

# 4. The Internet Architecture Board on Competing Roots

The Internet Architecture Board (IAB) describes itself as a "technical advisory group" of the Internet Society. It is a largely self-selected committee of 13 people, 6 of whom are nominated by a committee drawn from the Internet Engineering Task Force (these nominations must be approved by the Internet Society Board of Trustees), and another 6 basically recruited and selected by existing IAB members. The Chair of the IETF is an ex-officio member the IAB. Its meetings are also attended by the ICANN employee who performs the so-called Internet Assigned Numbers Authority (IANA) function. Because of their roots in the old DARPA-funded Internet technical community, IAB has very close ties to ICANN.

RFC 2826, The Internet Architecture Board's "Technical Comment on the Unique DNS Root," [IAB, 2000] is often cited in the policy debates over alternate roots as if it were the last word on the subject. According to RFC 2826, the DNS protocol was designed with the assumption that there would be only one authoritative root zone file. The statement goes on to describe some of the compatibility problems that might occur if computers attempting to resolve domain names are confused about the contents of the root zone file. The fundamental conclusion of RFC 2826 is this:

> …a degree of cooperation and agreed technical rules are required in order to guarantee the uniqueness of names. In the DNS, these rules are established independently for each part of the naming hierarchy, and the root domain is no exception. Thus, there must be a generally agreed single set of rules for [assigning the top-level domain names listed in] the root.

The statement is useful as a description of one of the original design parameters of the DNS. But RFC 2826 provides no guidance as to how to resolve the policy issues posed by the existence of competing or multiple roots.

## Alternate Roots Exist

The IAB statement was first issued in 1999 and became an "official" RFC in 2000. The document emerged from the heated policy controversies surrounding the early days of ICANN. While labeled a "technical comment," the document is most commonly cited in policy debates concerning competing roots, and tends to be used by ICANN and IETF as a technical "proof" that certain policies regarding the DNS root are undesirable.

In straining to clothe policy conclusions in the garb of technical expertise, however, the statement sacrifices precision. It states "it is not technically feasible for there to be more than one root in the public DNS." It is a rather strange claim. There *are* different root server systems in operation. These alternate root systems use the same DNS protocol and the same software implementation (BIND) as the ICANN root servers. Most, if not all, of them are capable of resolving all names under the IANA-delegated top-level domains. So it cannot be argued that they are not an implementation of the Domain Name System protocol. Nor it is technically correct to say that they are "private" rather than "public" name spaces. All of the alternate root systems are open to any ISP or end user who wishes to point resolvers or name servers in their direction.

Even if alternate roots did not exist now, nothing in the DNS protocol prevents a subset of the world's Internet service providers or end users from redirecting their name servers to some place other than the ICANN-administered root, if they wished to do so, at any time. The technical specifications for the DNS protocol do not say who administers the root, or name the IP addresses of root servers. Taken literally, the IAB statement is wrong: it *is* possible to have "more than one root in the public DNS." Indeed, it would be just as accurate for the technical comment to note that "nothing in the original design of DNS prevents the world's name server operators from disagreeing about who is or should be the authoritative root."

All the IAB is really capable of saying is that multiple roots may lead to compatibility problems in resolving names, as described under Type 2) and 3) competition above. But whether the value added by competing roots is worth the price of that incompatibility is a policy and economics question, not a technical issue.

More fundamentally, RFC 2826 provides no guidance as to how to respond to the compatibility problems posed by alternate roots once they exist. Should competing roots interconnect with each other? Should an incumbent root recognize the TLD assignments made by alternate roots and avoid adding new TLDs that conflict with them? Or should it maintain exclusivity and actively conflict with a competing root in order to drive it out of business? These are the real policy issues posed by root competition, and the IAB statement simply does not deal with them. Indeed, in many ways RFC 2826's insistence on the singularity of the root has become an obstacle to rational discussion of the problem.

## RFC 2826 and DNS Policy

RFC 2826 confuses three distinct questions: 1) the factual issue of whether multiple roots *can possibly exist*; 2) the normative questions whether alternate roots *should exist;* i.e., whether the *risks and costs of incompatibility are less desirable than the value created*; and 3) the practical question of *what is the appropriate policy response to alternate roots*, once they do exist and a normative determination has been made.

If one believes that alternate roots *should not* exist, one must come up with concrete, feasible, and legal proposals to preserve the singularity of the root, and one must deal with a variety of socio-economic tradeoffs. One does not, for example, contribute much to crime control policy by repeatedly asserting that "crime is bad" or that "crime should not exist." An effective crime prevention policy must recognize the existence of crime and identify specific measures to reduce or eliminate it. Those measures must balance the goal of crime prevention against other social values, such as privacy, individual rights, and budget constraints.

Assuming one accepts the basic premise of RFC 2826 that the root should be coordinated, and one accepts the fact that alternate roots can and do exist, one can identify several different policy options that might be advanced to achieve coordination:

- ICANN could agree to coordinate its top-level domain assignments with those of alternate roots, by adding their TLDs to its root zone (i.e., move to Type 1 competition)
- The ICANN root could ignore alternate roots, but avoid assigning TLDs that conflicted with TLDs in alternate roots
- ICANN could assume that the alternate roots have no right to exist and assign TLDs that conflicted with TLDs in alternate roots in an attempt to drive them out of existence
- The backers of ICANN could seek legislation to ban alternate roots, or redefine the DNS protocols in ways that rendered any alternate root incompatible with the ICANN root.
- ICANN could set thresholds that determined when a TLD from another root would be entered into its own root, or fix some other algorithm determining its relationship with them.

RFC 2826 provides no guidance as to which of these options is best. Nor should it; the IAB's expertise is in protocols and Internet architecture, not public policy.

In sum, RFC 2826's statement that "there must be a generally agreed single set of rules for the root" is merely the starting point for policy discussion. The assertion is both uncontroversial and not dispositive of the policy options available. Advocates of a "single authoritative root" need to face the reality that portions of the Internet community can and are defecting from or supplementing the ICANN root. Asserting that a particular root server system "should be" authoritative and singular does not make it so. One can agree on the need for coordination at the root level without necessarily agreeing that ICANN is the sole or proper source of those rules. Nor does the general need for a single set of rules eliminate the legitimacy and benefit of debate over what those rules should be.

## 5. What Sustains Alternate Roots?

The Internet technical community tends to view competing or alternative roots as a form of heresy. Those who advocate it are dismissed as anarchic troublemakers who want to undermine order for its own sake, or as confused people who don't understand DNS technology.

There is a simpler and more accurate explanation for the existence of alternate roots. Alternate roots arise primarily for economic reasons. They reflect a serious disequilibrium between the demand for new top-level domain names and the willingness of the root administrator to supply them. That conclusion is supported by an analysis of the historical origin and recent growth of alternate root efforts, and from the specific form that root-competition has taken.

## The Rise and Fall and Rise of Alt.Roots

The first organized effort to form an alternate DNS root emerged in 1996. This occurred shortly after Network Solutions, Inc. (NSI) began charging $50 per year for domain name registrations. It quickly became clear that in the environment of explosive growth created by the rise of the World Wide Web, domain name registration in generic top-level domains could be a lucrative business.

The Internet community, still organized informally around US Government-funded education and research networking initiatives, tried to coordinate efforts to create many new top-level domains in late 1995. There was widespread agreement that competition with Network Solutions was necessary and desirable. These efforts, loosely presided over by the late Jon Postel, produced an Internet draft proposing to create hundreds of new top-level domains. [Postel, 1996]. In anticipation of the award of new TLDs, some entrepreneurs, such as Eugene Kashpureff, Karl Denniger, Chris Ambler and Simon Higgs, began to operate "experimental" registries in 1996. Based on first occupation and use, they asserted exclusive claims on particular TLD strings, such as *.biz*, *.web*. or *.news*. Rancorous differences emerged, however, between these entrepreneurs and the traditional DARPA and NSF-funded Internet engineers led by Postel. The disagreements centered on the payment of fees for TLDs and the legitimacy of the decision making process. Other interests, notably the ITU and the trademark interests, also began to question the legitimacy of Postel's plan. Impatient with the delays and confusion surrounding the introduction of new TLDs, the entrepreneurs began to operate what they called "alternate roots" and to explicitly and publicly challenge the need for a single, authoritative root run by Postel. In 1997 another entrepreneur, Paul Garrin, set up an alternate name space with hundreds of generic top-level names (known as Name.Space) to promote the concept of non-exclusive generic TLDs. In this, the first round of root competition, the motivation for root-competition was the public's interest in registering names under more descriptive TLDs, and the inability of the Internet community to respond in a timely fashion to that market demand.

The alternate roots failed to achieve critical mass, however. The inherent handicap created by network effects, their lack of financing and sound business practices, and their inability to work together – symbolized most dramatically by the breakup of the eDNS coalition – led to their marginalization. The creation of ICANN from mid-1997 to the end of 1998 diverted attention from alternate roots even more, as those who wanted to operate or use new TLDs hoped that either the US Government's NTIA or the new private governance entity would formally authorize them. By the middle of the year 2000, only an estimated 0.3% of the world's name servers pointed to alternate roots. (Baptista, 2000)

However, ICANN's efforts to create new top-level domains took a long time. ICANN did not solicit bids for new TLDs until August of 2000, nearly two years after its creation. It required bidders for new TLDs to provide a $50,000 non-refundable fee to finance its evaluation and implementation. (Ironically, Postel's controversial 1996 proposal would have charged only $10,000 to *successful* registry applicants, yet led to the protests that sparked the first alternate root movement. ICANN was proposing to make organizations pay five times as much just to be *considered*.) Despite the slow pace and the costly application process, ICANN received applications from 43 different organizations seeking nearly 300 top-level names. Some of the bidders were operators of alternate roots, including Ambler's *.web* registry, and Paul Garrin's Name.space.

In November 2000, ICANN finally authorized only seven new top-level domains. Only three of the new ones (*.biz*, *.info*, and *.name*) were targeted at larger commercial markets. The rest were sponsored or restricted domains targeted at specific communities, such as museums, cooperatives, and the airline industry. The successful bidders, moreover, were almost all closely tied to ICANN or the Internet Society. Network Solutions, which already controlled *.com*, *.org* and *.net*, and Register.com, the second-largest registrar after Network Solutions, formed a consortium called Afilias with several other registrars to successfully bid for *.info*. Register.com also received the TLD *.pro*. CORE, a group of registrars that played a major political role in supporting and creating ICANN, received the contract for operating the *.museum* TLD and was also a member of the Afilias group. Thus, not only was the number of new TLDs small, but the awards went to a very limited group of market participants, to the exclusion of many viable new entrants.

An important and interesting aspect of ICANN's new TLD decision involved its relationship to some of the remaining alternate root operators. The Afilias consortium's first choice for a TLD was *.web*. At the November meeting, Board member Vinton Cerf opposed awarding *.web* to the Afilias consortium, explicitly acknowledging Image Online Design's prior claim and use of the *.web* TLD in an alternate root. Cerf's motion was opposed by Board member Hans Kraijenbrink, who explicitly acknowledged the conflict and asserted that ICANN ought not to recognize any claims to TLDs that emerged from outside of its own process. In a tense and close vote, Cerf and the ICANN Board "liberals" prevailed over the conservatives. Afilias was awarded *.info* instead of *.web*, and IOD was encouraged to wait for the second round.[4] On the other hand, one of the other successful bidders, a joint venture between Melbourne IT and NeuStar, applied to operate the *.biz* top-level domain. Biz was one of the original top-level domains operated by alternate root leader Karl Denninger in 1996. Denninger abandoned the TLD in 1998, but in the Spring of 2000, well before ICANN's call for bids, a businesswoman named Leah Gallegos took over the .biz domain and began operating it from within an alternate root system. [Gallegos, 2001] Unlike the *.web* case, Gallegos's prior claim was not recognized by ICANN, and *.biz* was awarded to the NeuStar/Melbourne IT joint venture.

ICANN's long-awaited decision on new TLDs, then, had three key features from a policy and economic standpoint. First, it introduced a very small group of TLDs, despite the existence of viable applications for many more; second, it limited the choice awards to a small group of businesses that were already privileged players in ICANN's process; third, it sent out mixed signals about the relationship between the ICANN root and competing roots. In the *.web* decision, it avoided conflict with an alternate root; in the *.biz* decision, it did not.

---

[4] These conversations are recorded and available for public viewing on the Harvard Berkman Center's web site. (http://cyber.law.harvard.edu/scripts/rammaker.asp?s=cyber&dir=icann&file=icann-111600&start=6-16-00)

Predictably, the artificial scarcity maintained by ICANN's TLD awards reinvigorated the alternate root movement. In the Spring of 2001 New.net, a company backed by substantial amount of venture capital, created 20 new top-level domains. The entrepreneurs behind New.net were unsuccessful bidders in the first round of ICANN's new top-level domain process. New.net's challenge was joined by the *.kids* registry, another unsuccessful bidder in the first round. End users could use or see the New.net domains by installing a software plug-in to their browsers. New.net also formed alliances with mid-sized Internet service providers to support the new domains by operating an alternate root. New.net does seem to be coordinating with the Open Root Server Confederation (ORSC), but not with the Name.space alternate root system (which was incompatible with ORSC).

This paper is being presented almost exactly a year since ICANN's decision to authorize the 7 new top-level domains. Only two of them are operational now, and those two only began to accept "sunrise" or "landrush" registrations in late August. The other new TLD delegees are bogged down in difficult contractual negotiations with ICANN, and complain frequently of ICANN's efforts to "micromanage" their affairs. Moreover, ICANN treated the rollout of new TLDs as an "experiment" that would be carefully studied and set no schedule for adding new top-level domains.

In contrast, New.net has aggressively expanded the name space, adding many new names to its first batch of 20, including names in foreign languages. It is clear that to the extent there is a viable market for new top-level domain names, New.net is responding to it more quickly and efficiently than the ICANN process, despite the tremendous handicap of not being visible in the ICANN root. [New.net Policy Document, 2001].

In short, the rise and fall of alternate root activity tracks very closely the efforts of the established root to add new top-level domains. When the process of adding new TLDs to the incumbent root is stymied or in disarray, alternate roots grow. When there appears to be hope that the process will accept new TLDs, alternate roots wither.

The relationship is clinched by the following facts:
- Alternate root competition has invariably taken the form of Type 2 competition. That is, competitors have attempted to add new TLDs to the set supported by ICANN. None of them have made assignments that conflict with ICANN TLDs. It is apparent, therefore, that the primary motive of these efforts is to find a way to add TLDs to the available set.
- Prior to the creation of ICANN, Name.Space repeatedly petitioned the US Government to enter its TLDs into the incumbent DNS root. Likewise, key backers of the alternate root movement were strong supporters of the US Government's Green Paper, which advocated rapid addition of 5 new TLDs into the established root, obviously in the hopes that they would be selected for that privilege.
- The major advocates and implementers of alternate roots – New.net, Image Online Design's .web registry, and Name.Space – all applied for TLDs via the ICANN process (and all were rebuffed). That is, all of them would strongly prefer to be visible in the ICANN root rather than attempting to establish an alternate root.

Evidence suggests very strongl that competition in the root zone is fueled primarily by restrictions in the supply of new TLDs.

## Conclusion

Why then does ICANN insist on maintaining artificial scarcity? The reason is rooted in the political constraints imposed on ICANN. As a corporatist "industry self-regulatory" body, ICANN represents a coalition of the Internet technical community, the intellectual property interests, incumbent registries, and a few major telecommunication and e-commerce firms. Few of these groups have anything to gain from adding new tlds; many have a direct economic or political interest in preventing it. More fundamentally, tight regulation of access to the root provides the leverage for the imposition of regulatory policy on the domain name industry. If ICANN were truly a "technical coordinator," as claimed by Ira Magaziner and the US Department of Commerce when it was created, it would passively accept and coordinate applications for new TLDs rather than actively restricting access and regulating their services and characteristics. Such a "thin ICANN" would be unable to impose on the DNS industry the kinds of trademark and market structure regulations that certain interests want. Nor would it provide the established technical hierarchy – the Jesuits of the Internet-Catholic faction – the degree of power over the Internet's technical evolution that they desire.

# Appendix: Time Line of ICANN's encounter with competing root policy

## November 2000

The ICANN Board authorizes 7 new TLDs. At that meeting, Board member Vinton Cerf opposes awarding the TLD string *.web* to the Afilias consortium, explicitly acknowledging Image Online Design's prior claim and use of the *.web* TLD in an alternate root. *Biz* assignment is made to Neulevel/Melbourne IT joint venture, despite prior claim of a small alternate root operator.

## February 2001

At US Congressional hearings, ICANN is grilled about its new TLD process. Vint Cerf defends the small number and apparently arbitrary selection process by claiming that the new TLDs were part of a "proof of concept" designed to test the feasibility of various types of new domains and admits that many viable proposals were not selected. Leah Gallegos, proprietor of the *.biz* top-level domain in the alternate root systems, testifies that she is being driven out of business by ICANN.

## March 2001

New.net, a company formed by a losing applicant in ICANN's new TLD lottery, along with another losing applicant, *.kids*, announce the creation of 20 new top-level domains supported by both a browser plug-in and an alternate root.

## April 2001

The DNSO General Assembly Chair calls for discussions of TLD name collisions, referring specifically to the *.biz* decision. The General Assembly also forms an email discussion list devoted to the topic of alternate roots.

## May 9 2001

The Names Council of the DNSO forms a Task Force on Multiple Roots and begins to assemble a "briefing paper" on the issues for discussion in Stockholm. Council is split on whether to form an open working group to take up the issue, or to avoid the issue.

## May 28 2001

ICANN CEO Stuart Lynn posts a "discussion draft" announcing that ICANN supports a "single, authoritative root." The document attacks the legitimacy and "insular" motives of the promoters of alternate roots.

## May 2001

For a brief period the US Federal Trade Commission posts a document on its web site warning consumers against "unauthorized top-level domains." The document is rumored to have been put there in response to pressure from the US Commerce Department and ICANN. A short time later, due to rumored lobbying pressure from New.net and Image Online Design, that language is removed from the document.

### June 3, 2001

At the Stockholm meeting, the DNSO Names Council votes 12-9 against forming a working group on the problem of multiple roots and instead requests "consultation" from the PSO and ASO. A few moments later it rescinds that vote and votes that "alternate roots are outside the scope of the DNSO."

### June 4, 2001

At the Stockholm meeting an open discussion of the Lynn draft is held before the Board. Supporters of the draft dominate the forum. At the meetings of the DNSO Business and ISP constituencies, New.net representatives receive harsh criticism.

### July 9, 2001

A slightly revised Discussion draft is issued by ICANN as an official policy, "Internet Coordination Policy #3." The policy was never formally debated or approved by either the DNSO or the ICANN Board. ICANN President Lynn justifies this by asserting that the document "reaffirms existing policy" rather than creating a new one.

### August 2001

Law professors and ICANNWatch editors Michael Froomkin and Jonathan Weinberg file a reconsideration request for ICP-3, arguing that ICP-3 bypassed DNSO's allegedly bottom-up policy processes. ICP-3's conclusions, they argue, "are not, even arguably, settled policy, 'developed through previous ICANN processes or received by ICANN at its creation.' Rather, they are new policies, announced in this document."